\documentclass[12pt,a4paper]{article}
\usepackage[dvips]{graphicx}
\usepackage{citesort}
\usepackage{amsmath,amssymb}

\makeatletter

\begin{document}

\begin{titlepage}

\vspace{.5cm}

\begin{center}
{\Large \bf Final State Interactions in the 
$D_s^+ \to \omega \pi^+$ and $D_s^+ \to \rho^0 \pi^+$ Decays\\} 

\vspace{.5cm}

{\large \bf S. Fajfer$^{a}$, A. Prapotnik$^{b}$, 
P. Singer$^{c}$ and J. Zupan$^{b,c}$\\}

\vspace{0.5cm}
{\it a) Department of Physics, University of Ljubljana, 
Jadranska 19, 1000 Ljubljana, Slovenia}

{\it b) J. Stefan Institute, Jamova 39, P. O. Box 300, 1001 Ljubljana, 
Slovenia}\vspace{.5cm}

{\it c) Department of Physics, Technion - Israel Institute  of Technology, 
Haifa 32000, Israel\\}

\end{center}

\vspace{1.5cm}

\centerline{\large \bf ABSTRACT}

\vspace{0.5cm}
We investigate  the decay mechanisms in the
$D_s^+ \to \omega \pi^+$ and $D_s^+ \to \rho^0 \pi^+$ 
transitions. The naive factorization ansatz predicts vanishing amplitude 
for 
the $D_s^+ \to \omega \pi^+$ decay,  
while the $D_s^+ \to \rho^0 \pi^+$ 
decay amplitude does have an annihilation contribution also in this
limit.  Both decays 
can proceed through intermediate states of hidden strangeness, e.g. 
$K,K^*$, which we estimate in this paper. These contributions can 
explain the experimental value for the $D_s^+ \to \omega \pi^+$ 
decay rate, which no longer can be viewed as a clean signature of 
the  
annihilation decay of $D_s^+$. The combination of the $\pi(1300)$ 
pole dominated annihilation contribution and the  internal $K,K^*$
exchange can saturate present experimental upper bound on   
$D_s^+ \to \rho^0 \pi^+$ decay rate, which is therefore expected 
to be within the experimental reach.  Finally, the proposed mechanism
of hidden strangeness FSI constitutes only a small correction 
to the Cabibbo allowed decay rates
$D_s\to KK^{*}, \phi \pi$, which are well
described already in the factorization approximation.

\end{titlepage} 


\section{Introduction}
\begin{figure}
\begin{center}
\includegraphics[width=8cm]{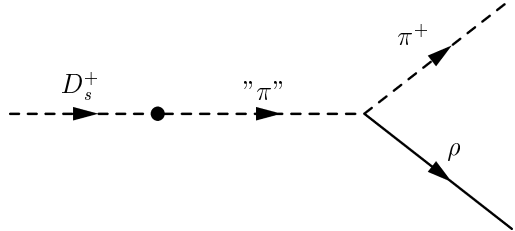}
\caption{\footnotesize Annihilation diagram of $D_s^+ \to \rho^0 \pi^+$
decay with $''\pi''=\pi, \pi(1300), \pi(1800)$, with dot representing the weak vertex.}\label{fig-annih}
\end{center}
\end{figure}

It has been suggested  \cite{balest}, that the observation of the $D^+_s\to \omega \pi^+$
decay
\begin{equation}
 BR(D_s^+ \to \omega \pi^+) = (2.8 \pm 1.1) \times
10^{-3} \label{omegapi}
\end{equation}
can be seen as a clean signature of the annihilation decay of 
$D_s^+$. The sizes of these contributions are of great
phenomenological interest, but  are very hard to obtain from
theoretical considerations.  In particular, the factorization approximation 
gives vanishing prediction for $BR(D_s^+ \to \omega \pi^+)$, as an
immediate consequence of the fact that due to the isospin and $G$ parity  
$I^{G}=1^+$ of $\omega \pi^+$ state  there are no annihilation contributions
with one intermediate resonant state (cf. also Fig. \ref{fig-annih}). The
interpretation of branching ratio \eqref{omegapi} as an (almost) purely
annihilation decay, 
would then also give important information about the,
otherwise relatively poorly known,  sizes
of annihilation diagrams in other related hadronic decays.  In this
paper, however, we will argue that the experimental value for the  
$D_s^+\to \omega \pi^+$ transition can be accommodated (already)  by
considering {\it only} color suppressed
spectator decay with subsequent final state interactions
(FSI). This leaves little room for unambiguous study of annihilation
effects from the $\omega \pi^+$ decay mode. 

Another issue that we address is the size of the $D_s^+ \to \rho^0
\pi^+$ transition. The experimental situation regarding this decay is
somewhat unclear.  The difficulties arise from the fact that the
$D_s^+ \to \rho^0\pi^+$ decay is observed as a resonance in the three
body decay $D_s^+ \to \pi^-\pi^+\pi^+$, which is dominated by the decays
through isoscalar resonances, $f_0(980)\pi^+$, $f_0(1370)\pi^+$ \cite{Frabetti:1997sx,aitala}. The
decay $D_s^+ \to \rho^0\pi^+$ therefore constitutes just a small
correction to these dominant processes. The two available experimental
analyses give
\begin{align}
 BR(D_s^+ \to \rho^0 \pi^+)& < 7 \times 10^{-4} \qquad\qquad\quad\cite{Frabetti:1997sx}\label{upper_bound}\,,\\
 BR(D_s^+ \to \rho^0 \pi^+)&= (5.9\pm4.6) \times 10^{-4} \quad\cite{aitala}\,,
\end{align}
where the first is a $90\%$ CL limit, while in the second case we have
added all the errors in quadrature. Clearly, the two
measurements are in agreement with each other. The error on the
number from \cite{aitala} is still very large and can at best be viewed
as an indication toward possible value for  $BR(D_s^+ \to \rho^0
\pi^+)$. In particular, it is still compatible with zero. As we will
show in the following, however, the expectations for  $BR(D_s^+ \to \rho^0
\pi^+)$ one gets from spectator decays with FSI and/or annihilation diagrams are just in the same
ballpark as the above experimental indications.

An important difference between $D^+_s\to \omega\pi^+$ and $D^+_s\to
\rho^0 \pi^+$ transitions
are the the quantum numbers of the final state. The $(\omega \pi^+)$ is a $I^G
(J^P)=1^+(0^-)$ state, while the $(\rho^0\pi^+)$ can be either in the
$I^G=1^-$ or $2^-$ state, again with $J^P=0^-$. A scan through PDG
book \cite{PDG} reveals, that there are no resonant states with $I^G
(J^P)=1^+(0^-)$
assignment, while there are three known states with $I^G
(J^P)=1^-(0^-)$, the
two strongly decaying resonances
$\pi(1300)$ and $\pi(1800)$, and the pion.

The presence of resonances near the $D_s$ meson mass with the quantum
numbers of $(\rho^0\pi^+)$ state indicates  an
enhancement of annihilation contributions, which in this case can
proceed through an intermediate resonating state (see Fig. \ref{fig-annih}). 
To gain an insight on the numerical importance of the annihilation
processes we estimate the contribution coming from the $\pi(1300)$
intermediate state. To do so, we need to estimate the decay constant
of $\pi(1300)$. Taking the PDG \cite{PDG} upper bound for $\tau \to
\pi(1300) \nu_\tau$ one arrives at $f_{\pi (1300)}< 4$ MeV. In the
factorization approximation for the weak vertex (for more details see
section \ref{FSI}) we then get 
\begin{equation}
BR(D_s^+ \to \rho^0\pi^+)_{\pi(1300)}< 7 \times 10^{-4}\,, \label{annihl}
\end{equation}
where we have used $f_{D_s}=230$ MeV, together with
the conservative assumptions of $BR(\pi(1300)\to \rho \pi)\sim 100\%$
and $\Gamma(\pi(1300))$ equal to its upper experimental bound of $600$
MeV.  Most probably this slightly overestimates the contribution of
$\pi(1300)$ intermediate resonant state to the $D_s^+ \to \rho^0\pi^+$
decay width, as also the presence (but not the size)  of other decay channels of $\pi(1300)$ has been
seen experimentally (e.g. $\pi(1300)\to (\pi \pi)_{S-{\rm wave}}
\pi$ \cite{PDG}). Note in particular that the above assumptions about the
$\pi(1300)\to \rho \pi$ decay width correspond to the case where
 $g_{\rho\pi\pi(1300)} \simeq g_{\rho\pi\pi}$. We therefore do not
expect the $\pi(1300)$ contribution to $BR(D_s^+ \to \rho^0\pi^+)$ to
lie significantly below the upper limit \eqref{annihl}.

The importance of the upper limit \eqref{annihl} is that
the contribution from $\pi(1300)$ can  even saturate the
90 \% CL experimental upper bound \eqref{upper_bound}. Of course the
actual size of $\pi(1300)$ contribution is not known and lies
somewhere below the upper bound \eqref{annihl}. Also, interference with other
annihilation contributions from intermediate $\pi$ and $\pi(1800)$ states can
somewhat change the above estimate (using PCAC, the contribution from $\pi$ was
found  in \cite{gourdin} to be negligible, while the contribution of 
$\pi(1800)$ is difficult to estimate due to the lack of experimental data). Furthermore, as we will
show in the next section, the contributions of final state
interactions fall in exactly the same range (cf. Eq. \eqref{FSI_rho} below). The lesson to be
learned from this simple exercise is that, unless there are large
cancellations, the value of  $BR(D_s^+ \to \rho^0\pi^+)$ is expected
to be near
to its present experimental upper bound and should be measured  
by the FOCUS collaboration in the near future \cite{pedrini}.

\begin{figure}
\begin{center}
\includegraphics[width=14cm]{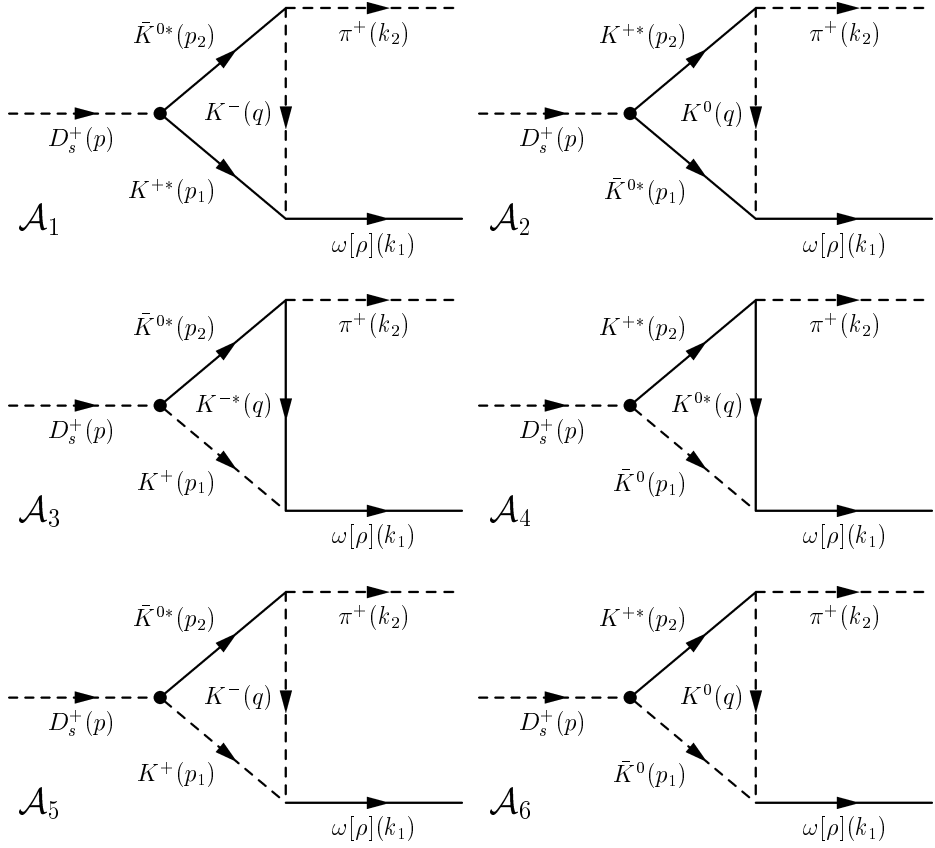}
\caption{\footnotesize{The $K, K^*$ meson contributions in the 
$D_s^+ \to \omega \pi^+$ and $D_s^+ \to \rho^0 \pi^+$ 
decay amplitudes.} }
\label{fig-diagrams}
\end{center}
\end{figure}

On the other hand note that, as explained above, {\it no such
resonance enhancements} of annihilation contributions are possible for
the $(\omega \pi^+)$ final state. How can then one explain a
relatively large experimental value for $BR(D_s\to \omega \pi^+)$,
Eq. \eqref{omegapi}? The answer lies in the fact that there are
multi-body intermediate states that do have correct values of $I^G$ and
$J^P$, for instance the two-body $(K^{(*)}\bar{K}^{(*)})$ states (see
Fig. \ref{fig-diagrams}). Moreover, such states of {\it hidden
strangeness} can be obtained from a (color suppressed) spectator decay
of the $D_s$ meson and are therefore expected to be sizable. We will
estimate the sizes of these contributions in the next section.

\begin{figure}
\begin{center}
\includegraphics[width=7cm]{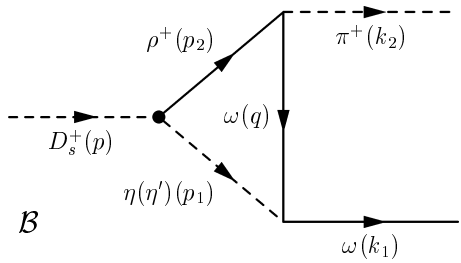}
\caption{\footnotesize{ The intermediate $\eta$, $\eta^\prime, \rho^+$ contributions 
in the $D_s^+ \to \omega \pi^+$ decay.}}\label{fig-eta}
\end{center}
\end{figure}

\section{Hidden strangeness FSI}\label{FSI}
In estimating the contributions from hidden strangeness intermediate
states that can arise from spectator quark diagrams, we will resort to the following simplifications
\begin{itemize}
\item Only two body intermediate states with $s, \bar{s}$ quantum
numbers will be taken into account. Moreover, only the contributions
of lowest lying pseudoscalar and vector states (neglecting their 
decay widths) will be considered.  Note that the re-scattering 
through intermediate $K, K^*$ states is possible for both 
$\rho^0 \pi^+$ as well as $\omega \pi^+$ final state (cf. Fig.
\ref{fig-diagrams}), while the re-scattering with intermediate $\eta$
or $\eta'$ is possible only in the case of $\omega \pi^+$ final state
due to isospin and $G$ parity conservation (cf. Fig. \ref{fig-eta}).

\item For the weak transition $D_s^+\to (K^{(*)}\bar{K}^{(*)})^+$ in the $D_s^+\to (K^{(*)}\bar{K}^{(*)})^+\to\rho^0\pi^+$ and $D_s^+\to (K^{(*)}\bar{K}^{(*)})^+\to
\omega\pi^+$ decay chains as well as for the weak transition $D_s^+\to
\eta(\eta')\rho^+$ in the  $D_s^+\to \eta(\eta')\rho^+\to
\omega\pi^+$ decay chain we will use the factorization approximation. The
weak Lagrangian is therefore
\begin{equation}
{\cal L}_{\rm weak}=-\frac{G_f}{\sqrt{2}} V_{cs} V_{ud}^*\big(a_1
(\bar{u}d)_H (\bar{s} c)_H +a_2
(\bar{s}d)_H (\bar{u} c)_H \big)\,,
\end{equation}
with $(\bar{u}d)_H, \dots$ the hadronized V-A weak currents, $  V_{ij}$ the CKM matrix elements and $a_{1,2}$ the effective (phenomenological)
Wilson coefficients taken to be $a_1=1.26$  and $a_2= -0.52$  \cite{bsw,buccel1a1,buccella2}. 

\item
Finally, the strong interactions are taken into account through the
following effective Lagrangian  \cite{chi1,chi2,chi3}:

\begin{equation}
{\cal L}_{\rm strong}=\frac{ig_{\rho\pi\pi}}{{\sqrt 2}}Tr(\rho^\mu[\Pi,\partial_\mu \Pi])-
4\frac{C_{VV\Pi}}{f}\epsilon^{\mu\nu\alpha\beta} Tr(\partial_\mu \rho_\nu \partial_\alpha \rho_\beta \Pi)\;,
\label{strong}
\end{equation}
where $\Pi$ and $\rho^\mu$ are $3 \times 3$ matrices
containing pseudoscalar and vector meson
operators respectively  and $f$ is a pseudoscalar decay constant 
(\ref{fconst}). We use numerical values $C_{VV \Pi}=0.33$, and $g_{\rho\pi\pi}= 5.9$
\cite{chi1,chi2,chi3}.

\end{itemize}

We
have checked that the use of factorization (at tree level, with values of form factors as given below in Eqs. \eqref{PvV}-\eqref{fconst} and in Table \ref{formfaktorji1}) for  the
$D_s^+ \rightarrow K^{*+} \bar K^{*0}$, $D_s^+ \rightarrow K^+ \bar K^{*0}$ 
and $D_s^+  \rightarrow \bar K^0 K^{*+}$  decays gives reasonable
estimates of the measured rates (note that we do not need 
$D_s^+ \rightarrow \bar K^0 K^{+}$ in further considerations), with
the results listed in Table \ref{Table-check}. In these results the
annihilation contributions have been neglected since they are an order
of magnitude smaller. We also neglect the triangle graphs of the sort shown on Figs. \ref{fig-diagrams}, \ref{fig-eta} as we expect them (based on the numerical results for $\omega \pi^+$, $\rho^0\pi^+$) to be suppressed relative to the tree level contributions.

\begin{table}
\begin{center}
\begin{tabular}{|c|c|c|c|} \hline
decay & $BR_{exp}$   & $BR_{th} $  \\ \hline
$D_s^+\to K^{*+}\bar K^{0*}$ & $(5.8 \pm 2.5)$\% & 6.4\%  \\ \hline
$D_s^+ \to K^{+}\bar K^{0*}$ & $(3.3 \pm 0.9)$\% & 3.4\%  \\ \hline
$D_s^+ \to K^{*+}\bar K^{0}$ & $(4.3 \pm 1.4)$\% & 2.7\%  \\ \hline
$D_s^+ \to \eta \rho^+$      & $(10.8 \pm 3.1)$\% & 5.6\%  \\ \hline
$D_s^+ \to \eta' \rho^+$     & $(10.1 \pm 2.8)$\% & 2.2\%  \\ \hline
\end{tabular}
\caption{\footnotesize{ The  branching ratios for the $D_s^+$ two body decays: in the second
column the experimental results are listed, while in the third the predictions
are given when the factorization approximation is used.}}
\label{Table-check}
\end{center}
\end{table}

The situation in the case of $\eta,\eta'$ intermediate states is not
so favorable. To treat the $\eta,\eta'$ mixing we use the approach of ref. \cite{kroll1} with the value of the
 mixing angle transforming between $\eta, \eta'$ and $\eta_q\sim(u\bar{u}+d\bar{d})/\sqrt{2}, \eta_s\sim s\bar{s}$ 
states taken to be $\phi=40^\circ$.  The factorization approach then
 gives a reasonable description of $D_s^+ \to \rho^+ \eta$ decay,
while it does not reproduce
 satisfactorily  the experimental result for $D_s^+ \to \rho^+ \eta'$
 (cf. Table \ref{Table-check}). This is
a known problem as $D_s^+ \to \rho^+ \eta'$ rate is  very difficult to
reproduce by any of present approaches \cite{buccel1a1,buccella2,hin}.
This inevitably introduces some further uncertainty into our approach,
yet the resulting uncertainty is not 
expected to affect significantly our main conclusions.

For the  weak current matrix elements between $D_s$ and vector or
pseudoscalar final states we use a common decomposition 

\begin{align}
\begin{split}
\langle V(k)|\bar q\Gamma^\mu
q|D_s(p)\rangle&=\epsilon^{\mu\nu\alpha\beta}\varepsilon_\nu
p_\alpha k_\beta \frac{2V(q^2)}{m_P+m_V}+2im_V\frac{\varepsilon \cdot
q}{q^2}q^\mu A_0(q^2)\\
+i(M+m_V)\Big[\varepsilon^\mu-&\frac{\varepsilon
\cdot
q}{q^2}q^\mu\Big]A_1(q^2)-i\frac{\varepsilon\cdot q}{M+m_V}\Big[P^\mu-\frac{M^2-m_V^2}{q^2}q^\mu\Big]
A_2(q^2)\;,
\label{PvV}
\end{split}
\\
\langle P(k)|\bar q\Gamma^\mu q|D_s(p)\rangle&=
\left(P^\mu-\frac{ (M^2-m^2)}{q^2}q^\mu\right)F_+(q^2)+
\frac{(M^2-m^2)}{q^2} q^\mu F_0(q^2)\;,
\end{align}
where $V$ is a vector meson characterized by the polarization vector
$\varepsilon^\mu$ and mass $m_V$, while pseudoscalar mesons $P,D_s$
have masses $m, M$. We use further abbreviations
$\Gamma^\mu=\gamma^\mu(1-\gamma^5)$, $q^\mu=p^\mu-k^\mu$, and
$P^\mu=p^\mu+k^\mu$.

For the $q^2$ dependence of the form factors
we use results of \cite{FF}, based on a quark model calculation
combined with a fit to lattice and experimental data. Ref. \cite{FF}
provides a simple fit to their numerical results with the form factors
$F_+(q^2)$, $V(q^2)$ and $A_0(q^2)$ described by double pole $q^2$ dependence
\begin{equation}
f(q^2)=\frac{f(0)}{(1-q^2/M^2)(1-\sigma q^2/M^2)}\;,
\label{double}
\end{equation}
while single pole parameterization
\begin{equation}
f(q^2)=\frac{f(0)}{(1-\sigma q^2/M^2)}\;,
\label{single}
\end{equation}
can be  used for $A_{1,2} (q^2)$, as the contributing resonance lie
farther away from the physical region (note that this parameterization
applies also to $F_0$ form factor, which however does not contribute
in the processes we discuss in this paper). The values of $f(0)$ and
$\sigma$ are listed in Table \ref{formfaktorji1} and are taken from  \cite{FF}.
We use $M=1.97$ $ \rm GeV$ in the expression for $A_0$, while
$M=2.11$ $ \rm GeV$ is used for other  
form factors \cite{FF}.  Incidentally, the parameterizations of form
factors \eqref{double}, \eqref{single}
 make all the loop diagrams in Figs.
\ref{fig-diagrams} and \ref{fig-eta} finite, so that no regularization
procedure is needed (we have also checked that the numerical results   do not
change significantly, if one uses a single pole parameterization
instead of
\eqref{double} and then uses a cut-off regularization with some scale
above but close enough to $D_s$ meson mass) .

\begin{table}
\begin{center}
\begin{tabular}{|c|c|c|c|c|c|c|}
\hline
form factor  & $F_+$ & $V$  & $A_0$ & $A_1$ & $A_2$ & $F_{\eta_s,+}$ \\ \hline 
$f(0)$      & 0.72  & 1.04 & 0.67  & 0.57  & 0.42  &	0.78     
 \\ \hline
$\sigma$    & 0.2   & 0.24 & 0.2   & 0.29  & 0.58  &    0.23      \\ \hline
\end{tabular}
\label{formfaktorji1}
\caption{\footnotesize{Form factors at $q^2=0$  \cite {FF}. The results in the first five 
columns are for $D_s \to K,K^* l \nu_l$ transitions. 
The last column stands  for the form factor appearing in 
$D_s \to \eta_s l \nu_l$, 
(the $s \bar s$ component of $\eta, \eta'$) 
transition.}}
\end{center}
\end{table}
For the decay constants, defined through 
\begin{equation}
\langle 0|\bar q\gamma^\mu \gamma_5 q|P(p)\rangle=if_P p^\mu\;, \qquad 
\langle 0|\bar q\gamma^\mu q|V(p)\rangle=g_V \varepsilon^\mu\;.
\label{fconst}
\end{equation}
we use the following values:
 $f_D=0.207\; \rm{GeV}$ and $f_{Ds}=1.13 f_D$ as obtained on the
lattice \cite{fdlat} 
and for the rest  $f_K=0.16\;\rm{GeV}$, $|g{_K*}|=0.19\;\rm{GeV}^2$, $|g_\rho|=0.17\;\rm{GeV}^2$ 
and $|g_\omega|=0.15\;\rm{GeV}^2$ coming from the experimental measurements \cite{PDG}.

\begin{table}
\begin{center}
\begin{tabular}{|c|c|c|c|c|c|c||c|c|}
\hline 
$D_s^+ \to \omega \pi^+$& ${\cal A}_1$ & ${\cal A}_2$ & ${\cal A}_3$&
${\cal A}_4$ & ${\cal A}_5$& ${\cal A}_6$ & ${\cal B}_\eta$ &  ${\cal B}_{\eta^\prime}	$\\
\hline
${\cal A}_{iD}$ & $-0.7$ & $0.7$
&$-1.1$&$-1.4$&$11.3$&$12.5$&$1.3$&$3.6$\\
\hline
${\cal A}_{iA}$& $-0.7$ & $0.7$
&$3.3$&$1.5$&$-4.0$&$-19.7$&$-7.2$&$-3.7$\\
\hline
\end{tabular}\caption{\footnotesize{ The dispersive ${\cal A}_{iD}$  and absorptive ${\cal A}_{iA}$ 
parts of the amplitudes (in units of $10^{-3}$ ${\rm GeV}$)  for the 
$D_s^+ \to \omega \pi^+$ decay corresponding to the diagrams on
Fig. \ref{fig-diagrams} (${\cal A}_i$) and Fig. \ref{fig-eta} 
(${\cal B}_{\eta,\eta'}$).
 The
amplitudes for the $D_s^+ \to \rho^0 \pi^+$ decay (neglecting the mass
difference between $m_\rho$ and $m_\omega$) are obtained  by
inverting the sign of ${\cal A}_{iD}, {\cal A}_{iA}$ for even $i$,
while ${\cal B}_{\eta,\eta'}=0$.}}
\label{results}
\end{center}
\end{table}

The amplitudes for the $D^+_s \to \omega \pi^+$ and $D^+_s \to \rho^0 \pi^+$
decays can be written as:
\begin{align} 
{\cal A}(D^+_s(p) \to 
\omega(\varepsilon,k_1)\pi^+(k_2))&=\frac{G_f}{\sqrt{2}}\;\varepsilon
\cdot k_2\;\Big(\sum_i {\cal A}_i^{(\omega)}+{\cal B}\Big) \label{omega_amp}
, \\
{\cal A}(D^+_s(p) \to \rho^0(\varepsilon,k_1)\pi^+(k_2))&=\frac{G_f}{\sqrt{2}}\;\varepsilon \cdot k_2\;\sum_i {\cal A}_i^{(\rho)}\;\label{rho_amp}
,
\end{align}
with $\varepsilon$  the helicity zero polarization vector of the
$\omega$ or $\rho$ vector mesons, while  $k_2$ is the pion momentum. 
The reduced amplitudes ${\cal A}_i^{(\rho),(\omega)}$ and ${\cal B}$ correspond to the
diagrams on Figs. \ref{fig-diagrams} and \ref{fig-eta} respectively
(note that in our phase convention for Clebsch-Gordan coefficients
${\cal A}_i^{(\rho)}=(-1)^{i+1}{\cal A}_i^{(\omega)}$, i.e. the even
diagrams  for the two
channels on  Fig.  \ref{fig-diagrams} differ by a sign).
The explicit expressions  can be found in  Appendix. The
numerical values for  ${\cal A}_i^{(\rho),(\omega)}$ and ${\cal B}$ are given in Table \ref{results}.

Combining the above results we arrive at the  prediction  
\begin{equation}
BR(D_s^+ \to \omega \pi^+)=3.0
\times 10^{-3}.
\end{equation}
 Note that in this calculation we have used the
factorization approximation for the diagram of Fig.  \ref{fig-eta},
which as stated above, does not work well for $D_s^+ \to \rho^+ 
\eta,\eta'$ transition.\footnote{Adding hidden strangeness FSI  
in the
spirit of this paper to the $D_s^+\to \rho^+ \eta'$
channel does not mend the discrepancy, as it gives an order of
magnitude smaller contribution.}  If one instead uses experimental input to rescale
the corresponding amplitudes one ends up with 
$BR(D_s^+ \to \omega \pi^+)=4.4\times 10^{-3}$.
We see, that it is actually very easy to explain the experimental
result \eqref{omegapi} {\it entirely} in terms of spectator quark
transition together with FSI. The claim, that $D_s^+ \to \omega \pi^+$
can be used as a probe of annihilation contributions in hadronic
physics is therefore not justified.

As what concerns the  $D_s^+ \to \rho^0 \pi^+$ transition, the FSI
contributions alone
give
\begin{equation}
BR(D_s^+ \to \rho^0 \pi^+)_{\rm FSI}=0.7 \times 10^{-3} \,,\label{FSI_rho}
\end{equation}
which is almost exactly the same as our estimate of the upper bound
on  the annihilation
contribution \eqref{annihl} and actually coincides with the present
90\% CL upper bound. If there is no destructive interference
between these two contributions and the contributions of FSI through
higher resonances that we did not take into account,  this decay mode
should be established in the near future. This
expectation is supported also by other theoretical approaches which give the rate for 
$D_s^+ \to \rho^0 \pi^+$ to be equal 
\cite{buccella2} or even larger than 
the rate for $D_s^+ \to \omega \pi^+$ decay \cite{hin,rosner}. 

Possible
cancellation that {\it may} occur, however, make the theoretical
predictions rather uncertain. Adding the FSI contribution and the maximal
annihilation contributions \eqref{annihl} with alternating signs gives a fairly large
interval 
\begin{equation}
BR(D_s^+ \to \rho^0 \pi^+)=(0.05-3.5)\times 10^{-3}\,.
\end{equation}
We note that the experimental uncertainties translating in the input parameters can change the values
for $BR(D_s^+ \to \rho^0 \pi^+)$ and $BR(D_s^+ \to \omega \pi^+)$ by about 20\%. 

Finally, we mention that the kind of FSI contributions we were
considering in this paper will not be the leading contribution in  the $D_s^+\to
\phi \pi^+$ transition, namely, this decay can proceed through
spectator quark transition directly. Use of factorization approximation for the weak vertex
leads to a prediction $BR(D_s^+\to
\phi \pi^+)=4.0\%$, which is already in excellent agreement with the
experimental result of $3.6 \pm 0.9 \%$. Inclusion of FSI reduces the
theoretical prediction from 4\% to  $\sim 3.6\%$ and does not spoil
the agreement with the experiment (it actually even improves it). The
size of the shift also indicates
that FSI of the type described in the present paper are  in the case of $D_s^+\to
\phi \pi^+$ transition a second order effect. Note as well, that the
size of the FSI correction is in agreement with the predictions for
$BR(D_s^+\to \rho^0\pi^+)$ and $BR(D_s^+\to \omega \pi^+)$, which are
an order of magnitude smaller than $BR(D_s^+\to\phi \pi^+)$.

In conclusion, we found that the hidden strangeness final state
interactions are very important in understanding the 
$D^+_s \to \omega \pi^+$ and $D^+_s \to \rho^0 \pi^+$ decay mechanism. 
The $D^+_s \to \omega \pi^+$ amplitude can be explained fully by this
mechanism.  As for the $D^+_s \to \rho^0 \pi^+$ amplitude the predictions
we obtain lie in a fairly large range due to possible cancellation
between FSI and single pole contributions. There is a hope that  $D^+_s \to \rho^0 \pi^+$
will be measured  soon, hopefully  shedding  more light on our understanding
of the $D^+_s \to \rho^0 \pi^+$ decay mechanism. Finally, we note that
the  hidden strangeness FSI discussed in this paper
represents a second order effect, the inclusion of which does not spoil the good
agreement of factorization approximation obtained for $D_s\to \phi
\pi, K K^*$.

\vskip5mm
\noindent
{\bf {\large Acknowledgments}}\\[2mm]
S.F., A.P. and J.Z. are
supported in part by the Ministry of Education, Science and Sport of
the Republic of Slovenia.

\appendix
\section{Decay amplitudes}
 In this appendix we list expressions for the reduced
amplitudes ${\cal A}_i$
and ${\cal B}$ defined through Eqs. \eqref{omega_amp}-\eqref{rho_amp} and
Figs. \ref{fig-diagrams}, \ref{fig-eta}. 
We present explicitly the  ${\cal A}_i^{(\omega)}$  reduced amplitudes only for
the  $D^+_s \to \omega \pi^+$ decay. The corresponding expressions
${\cal A}_i^{(\rho)}$ for the $D^+_s \to \rho^0 \pi^+$
decay are then obtained by inverting the signs of ${\cal A}_i^{(\omega)}$ for
even $i$:
\begin{align}
\begin{split}
\varepsilon \cdot k_2\,{\cal A}_1^{(\omega)}&
=\int\frac{d^4q}{(2\pi)^4}
a_2 g_K\frac{2}{m_*+M}\frac{V(0)}{1-p_2^2/M_{F}^2}
\epsilon_{\mu\nu\alpha\beta}p_2^{\alpha} p_1^\beta
\frac{-i(g^{\mu\mu^\prime}-p_2^\mu p_2^{\mu^\prime}/m_*^2)}{p_2^2-m_*^2}
\times \\ 
&\frac{-i(g^{\nu\nu^\prime}-p_1^\nu p_1^{\nu^\prime}/m_*^2)}{p_1^2-m_*^2}
\frac{i}{q^2-m^2}\frac{ig_{\rho\pi\pi}}{\sqrt{2}}(k_2-q)_{\mu^\prime}
\frac{i4C_{VV\Pi}\,\epsilon_{\nu^\prime\sigma\tau\delta}\,
p_1^\sigma\varepsilon^{\tau}k_1^\delta}{f_K\sqrt{2}}\;,
\label{A1}
\end{split}
\end{align}
\begin{align}
\begin{split}
\varepsilon \cdot k_2\,{\cal A}_2^{(\omega)}&=
\int\frac{d^4q}{(2\pi)^4}
a_2 g_K \frac{2}{m_*+M}\frac{V(0)}{1-p_1^2/M_{F}^2}
\epsilon_{\mu\nu\alpha\beta}p_2^{\alpha} p_1^\beta 
\frac{-i(g^{\mu\mu^\prime}-p_2^\mu p_2^{\mu^\prime}/m_*^2)}{p_2^2-m_*^2}
\times \\ 
&\frac{-i(g^{\nu\nu^\prime}-p_1^\nu p_1^{\nu^\prime}/m_*^2)}{p_1^2-m_*^2}
\frac{i}{q^2-m^2}\frac{ig_{\rho\pi\pi}}{\sqrt{2}}(q-k_2)_{\mu^\prime}
\frac{i4C_{VV\Pi}\,\epsilon_{\nu^\prime\sigma\tau\delta}\,
p_1^\sigma\varepsilon^{\tau}k_1^\delta}{f_K\sqrt{2}}\;,
\label{A2}
\end{split}
\\
\begin{split}
\varepsilon \cdot k_2\,{\cal A}_3^{(\omega)}&=
\int\frac{d^4q}{(2\pi)^4}a_2 g_K \frac{F_+(0)}{1-p_2^2/M_{F}^2} 
(p+p_1)_\mu
\frac{-i(g^{\mu\mu^\prime}-p_2^\mu p_2^{\mu^\prime}/m_*^2)}{p_2^2-m_*^2}
\frac{i}{p_1^2-m^2} 
\times \\ 
&\frac{-i(g^{\nu\nu^\prime}-q^\nu q^{\nu^\prime}/m_*^2)}{q^2-m_*^2}
\frac{i4C_{VV\Pi}\,
\epsilon_{\mu^\prime\alpha\nu^\prime\beta}\,p_2^\alpha q^\beta}{f_\pi}
\frac{i4C_{VV\Pi}\,
\epsilon_{\nu\sigma\tau\delta}\,q^\sigma\varepsilon^\tau k_1^\delta}
{f_K\sqrt{2}}\;,
\label{A3}
\end{split}
\\
\begin{split}
\varepsilon \cdot k_2\,{\cal A}_4^{(\omega)}&=
\int\frac{d^4q}{(2\pi)^4}a_2
\left(2im_*\frac{p_{1\mu} p_1^\lambda}{p_1^2}\frac{A_0(0)}{1-p_1^2/
M_A^2}
-i\frac{p_{1\mu}}{M+m_*}\frac{A_2(0)}{1-p_1^2/M_{F}^2}\right.
\times \\ 
&\left.
\left((p+p_2)^\lambda 
-\frac{M^2-m_*^2}{p_1^2}p_1^\lambda \right) \right)
(-if_K p_1)_\lambda 
\frac{i}{p_1^2-m^2} 
\frac{-i(g^{\mu\mu^\prime}-p_2^\mu
p_2^{\mu^\prime}/m_*^2)}{p_2^2-m_*^2}
\times \\ 
&\frac{-i(g^{\nu\nu^\prime}-q^\nu q^{\nu^\prime}/m_*^2)}{q^2-m_*^2}
\frac{i4C_{VV\Pi}\,
\epsilon_{\mu^\prime\alpha\nu^\prime\beta}\,p_2^\alpha q^\beta}{f_\pi}
\frac{i4C_{VV\Pi}\,
\epsilon_{\nu\sigma\tau\delta}\,q^\sigma\varepsilon^\tau k_1^\delta}
{\sqrt{2}f_K}\;,
\label{A4}
\end{split}
\\
\begin{split}
\varepsilon \cdot k_2\,{\cal A}_5^{(\omega)}&=
\int\frac{d^4q}{(2\pi)^4}a_2 g_K \frac{F_+(0)}{1-p_2^2/M_{F}^2} 
(p+p_1)_\mu
\frac{-i(g^{\mu\nu}-p_2^\mu p_2^{\nu}/m_*^2)}{p_2^2-m_*^2}
\times \\ 
&\frac{i}{p_1^2-m^2} \frac{i}{q^2-m^2}
\frac{ig_{\rho\pi\pi}(k_2-q)_\nu}{\sqrt{2}}
\frac{ig_{\rho\pi\pi}(q-p_1)\cdot \varepsilon}{2}\;,
\label{A5}
\end{split}
\end{align}
\begin{align}
\begin{split}
\varepsilon \cdot k_2\,{\cal A}_6^{(\omega)}&=
\int\frac{d^4q}{(2\pi)^4}a_2
\left(2im_*\frac{p_{1\mu} p_1^\lambda}{p_1^2}\frac{A_0(0)}{1-p_1^2/M_{A}^2}
-i\frac{p_{1\mu}}{M+m_*}\frac{A_2(0)}{1-p_1^2/M_F^2} \right.
\times \\ 
&\left. 
\left((p+p_2)^\lambda 
-\frac{M^2-m_*^2}{p_1^2}p_1^\lambda\right)
\right)  
(-if_K p_1)_\lambda  
\frac{-i(g^{\mu\mu^\prime}-p_2^\mu
p_2^{\mu^\prime}/m_*^2)}{p_2^2-m_*^2}
\times \\ 
&\frac{i}{p_1^2-m^2} \frac{i}{q^2-m^2}
\frac{ig_{\rho\pi\pi}(q-k_2)_\nu}{\sqrt{2}}
\frac{ig_{\rho\pi\pi}(q-p_1) \cdot \varepsilon}{2\sqrt{2}}\;.
\label{A6}
\end{split}
\end{align}
%
In expressions above  we used $m, m_*, M$ for the $K, K^*, D_s$ mass
respectively. The pole masses used in the form
factors are $M_F=2.11$ GeV and $M_A=1.97$ GeV.
The definitions of momenta $k_{1,2}$, $p_{1,2}$ and $q$  
are the same as the ones used on Figs.
\ref{fig-diagrams}, \ref{fig-eta}.

\noindent The ${\cal B}$ part of the $D_s^+ \to \omega \pi^+$ amplitude
(Fig. \ref{fig-eta}) is 
\begin{equation}\label{B_expl}
\begin{split}
\varepsilon \cdot k_2\,{\cal B }= 
\int\frac{d^4q}{(2\pi)^4}&a_1 g_\rho \frac{F_{+,\eta^{(\prime)}}(0)}{1-p_2^2/M_F^2} 
(p+p_1)_\mu
\frac{-i(g^{\mu\mu^\prime}-p_2^\mu p_2^{\mu^\prime})}{p_2^2-m_\rho^2}
\frac{i}{p_1^2-m_{\eta^(\prime)}^2} \times
\\
&\times\frac{-i(g^{\nu\nu^\prime}-q^\nu q^{\nu^\prime}/m_\omega^2)}{q^2-m_\omega^2}
\frac{4 i}{f_\pi}\sqrt{2}C_{VV\Pi}\,
\epsilon_{\mu^\prime\alpha\nu^\prime\beta}\,p_2^\alpha q^\beta\times\\
&\times\frac{4i}{f_{\eta^{(\prime)}}}K_{\eta^{(\prime)}}C_{VV\Pi}\,
\epsilon_{\nu\sigma\tau\delta}\,q^\sigma\varepsilon^\tau k_1^\delta
\;,
\end{split}
\end{equation}
while ${\cal B}=0$  for $D_s^+ \to \rho^0 \pi^+$. In \eqref{B_expl} we used the abbreviations
\begin{equation}
K_{\eta}=\cos\phi/\sqrt{2} 
\qquad 
K_{\eta^\prime}=\sin\phi/\sqrt 2
\end{equation}
and
\begin{eqnarray}
F_{+,\eta}(0)= \sin{\phi}F_{\eta_s,+}(0) \qquad \nonumber
F_{+,\eta^\prime}(0)= \cos{\phi}F_{\eta_s,_+}(0).
\end{eqnarray} 
Finally, the decay rate for the $D^+_s \to \omega(\rho^0)\pi^+$ is given by
\begin{equation}
\begin{split}
\Gamma=\frac{G_f^2 k^3}{8 \pi
m_1^2} \left|\sum_i {\cal A}_i^{(\rho),(\omega)}\; +{\cal B}
\right|^2\,,
\end{split}
\end{equation}
with $M$ the $D_s$ mass, and $m_{1,2}$ the vector and the pseudoscalar
meson masses respectively, while
\begin{equation}
k= \frac{1}{2M}[(M^2-(m_1+m_2)^2)(M^2-(m_1-m_2)^2)]^{1/2}
\end{equation}
is the size of the three-momentum of the final particles in the $D_s$
rest frame.


\begin{thebibliography}{10}

\bibitem{balest} 
R.~Balest {\it et al.}  [CLEO Collaboration],
Phys.\ Rev.\ Lett.\  {\bf 79}, 1436 (1997)
[arXiv:hep-ex/9705006].

\bibitem{Frabetti:1997sx}
P.~L.~Frabetti {\it et al.}  [E687 Collaboration],
Phys.\ Lett.\ B {\bf 407}, 79 (1997).
\bibitem{aitala}
E.~M.~Aitala {\it et al.}  [E791 Collaboration],
Phys.\ Rev.\ Lett.\  {\bf 86}, 765 (2001)
[arXiv:hep-ex/0007027].
 
\bibitem{PDG}
K.~Hagiwara {\it et al.}  [Particle Data Group Collaboration],
Phys.\ Rev.\ D {\bf 66}, 010001 (2002).


\bibitem{gourdin} 
M.~Gourdin, Y.~Y.~Keum and X.~Y.~Pham,
Phys.\ Rev.\ D {\bf 53}, 3687 (1996)
[arXiv:hep-ph/9506225].


\bibitem{pedrini} D. Pedrini, FOCUS collaboration, private communication.
\bibitem{bsw} M.~Bauer, B.~Stech and M.~Wirbel,
Z.\ Phys.\ C {\bf 34}, 103 (1987).

\bibitem{buccel1a1} 
F.~Buccella, M.~Lusignoli, G.~Miele, A.~Pugliese and P.~Santorelli,
Phys.\ Rev.\ D {\bf 51}, 3478 (1995)
[arXiv:hep-ph/9411286].

\bibitem{buccella2}
F.~Buccella, M.~Lusignoli and A.~Pugliese,
Phys.\ Lett.\ B {\bf 379}, 249 (1996)
[arXiv:hep-ph/9601343].

\bibitem{chi1} 
M.~Bando, T.~Kugo, S.~Uehara, K.~Yamawaki and T.~Yanagida,
Phys.\ Rev.\ Lett.\  {\bf 54}, 1215 (1985);
M.~Bando, T.~Kugo and K.~Yamawaki,
Nucl.\ Phys.\ B {\bf 259}, 493 (1985);
M.~Bando, T.~Kugo and K.~Yamawaki,
Phys.\ Rept.\  {\bf 164}, 217 (1988).


\bibitem{chi2} T.~Fujiwara, T.~Kugo, H.~Terao, S.~Uehara and K.~Yamawaki,
Prog.\ Theor.\ Phys.\  {\bf 73}, 926 (1985).

\bibitem{chi3} A.~Bramon, A.~Grau and G.~Pancheri,
Phys.\ Lett.\ B {\bf 344}, 240 (1995).

\bibitem{kroll1} T.~Feldmann, P.~Kroll and B.~Stech,
Phys.\ Rev.\ D {\bf 58}, 114006 (1998)
[arXiv:hep-ph/9802409];
T.~Feldmann, P.~Kroll and B.~Stech,
Phys.\ Lett.\ B {\bf 449}, 339 (1999)
[arXiv:hep-ph/9812269].

\bibitem{hin} I.~Hinchliffe and T.~A.~Kaeding,
Phys.\ Rev.\ D {\bf 54}, 914 (1996)
[arXiv:hep-ph/9502275].

\bibitem{FF} D.~Melikhov and B.~Stech,
Phys.\ Rev.\ D {\bf 62}, 014006 (2000)
[arXiv:hep-ph/0001113].

\bibitem{fdlat} 
D.~Becirevic,
Nucl.\ Phys.\ Proc.\ Suppl.\  {\bf 94}, 337 (2001)
[arXiv:hep-lat/0011075];
N.~Yamada,
arXiv:hep-lat/0210035;
S.~M.~Ryan,
Nucl.\ Phys.\ Proc.\ Suppl.\  {\bf 106}, 86 (2002)
[arXiv:hep-lat/0111010].



\bibitem{rosner} 
C.~W.~Chiang, Z.~Luo and J.~L.~Rosner,
Phys.\ Rev.\ D {\bf 67}, 014001 (2003)
[arXiv:hep-ph/0209272].


\end{thebibliography}
\end{document}